# LATTICE STRUCTURAL ANALYSIS ON SNIFFING TO DENIAL OF SERVICE ATTACKS

B.Prabadevi[1], N.Jeyanthi[1], Nur Izura Udzir[2] and Dhinaharan Nagamalai[3]

[1]School of Information Technology and Engineering, Vellore Institute of Technology, Vellore, India
[2]Department of Computer Science, Universiti Putra Malaysia, Malaysia
[3]Wireilla Net Solutions, Australia

*ABSTRACT*

*Sniffing is one of the most prominent causes for most of the attacks in the digitized computing environment. Through various packet analyzers or sniffers available free of cost, the network packets can be captured and analyzed. The sensitive information of the victim like user credentials, passwords, a PIN which is of more considerable interest to the assailants' can be stolen through sniffers. This is the primary reason for most of the variations of DDoS attacks in the network from a variety of its catalog of attacks. An effective and trusted framework for detecting and preventing these sniffing has greater significance in today's computing. A counter hack method to avoid data theft is to encrypt sensitive information. This paper provides an analysis of the most prominent sniffing attacks. Moreover, this is one of the most important strides to guarantee system security. Also, a Lattice structure has been derived to prove that sniffing is the prominent activity for DoS or DDoS attacks.*

## 1. INTRODUCTION

The most malicious case for DDoS [1] attacks is the sniffing which can be implemented by tools or can be done manually for attacking the victim system. The attacks caused by this process is called sniffing attacks and fully automated tool used is named as sniffer or packet analyzer. These sniffing attacks are one of the causes of DDoS attacks. Most of the sniffing attacks are involved in capturing sensitive information/data like user credentials, password, secret codes, PIN, and so on. Sniffer performs its operations in two modes, namely promiscuous and not-promiscuous. In the former one, the sniffer can pilfer information from all the devices connected to the network and in the later one, only the information going to and from its host system. Based on the type of information stealth and ways it performs the sniffing, it has broad classification as shown in Figure 1. Sniffing attack can be in two ways Active sniffing and passive sniffing. The former one can be detected where the attacker can sniff the traffic in the network, whereas the later one is difficult to detect but can be prevented [2]. Also, B. Prabadevi and N. Jeyanthi [2] present some of the active sniffing tools used to sniff the data via a network for launching various attacks in the network and anti sniffer tools.

Usually, these DDoS attacks can be caused by an individual or an attacker with the help of some handlers or agents/zombies responding to an attacker's command to exploit the victims. In the

International Journal of Computer Networks & Communications (IJCNC) Vol.11, No.4, July 2019

latter case, the attacker sends the command holding the victim's information and attack type to the handlers who will direct the zombies to propagate these camouflaged data packets to the victim's system. Here the victim can be a single host or a link termed as destination flooding and link flooding respectively. This process seems to be simple but causes a devasting effect without getting caught quickly, by making use of zombies through spoofing, and the sources are also distributed. This may either consume more bandwidth by exploiting the resources or crashes the entire system itself. This forms the basis for various forms of attacks on the internet. The vulnerabilities caused by DDoS attacks are growing tremendously from 171 in 1995, 7236 in 2007, 6058 in 2008, CERT investigated 130, 165 DDoS reports in 2013 and 2014. Especially in Lithuania CERT identified 2000 computers/ day were remotely accessed without owners' knowledge and 2500 computers/day in 2015 (CERT statistics) [3]. Akamai suggests DDoS attacked peaked at 300 GBPS in 2013and 400 Gbps in 2014, which had put 160,000 users offline for about multiple hours [4]. Most of the DDoS attacks go unreported.

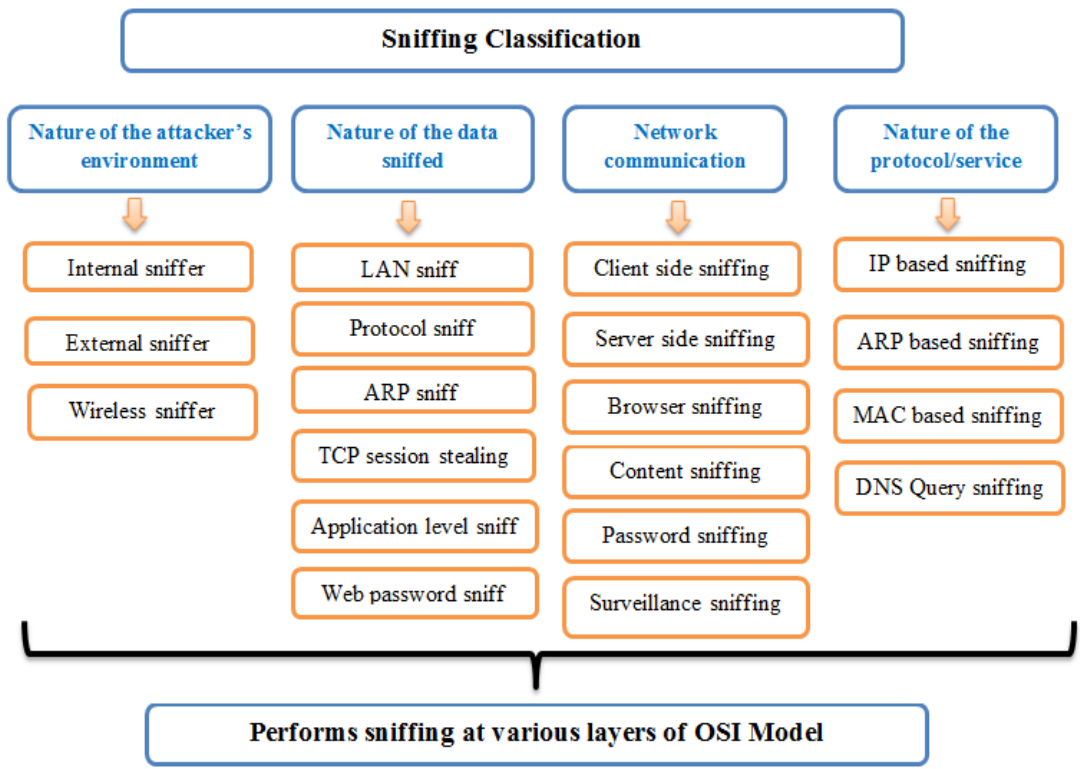

Figure 1. Classification of sniffing based on various scenarios

An adequate infrastructure with different levels of encryption must be provided, to overcome these attacks. The detection of these attacks remains to be challenging one than prevention. Most of the computing technologies hold both sensitive and non-sensitive data. Armor for segregating sensitive information from non-sensitive information with authentication at multiple levels is needed, which makes it difficult for the cyber antagonist to gain access to sensitive information. This helps the organization to overcome the issues with Bring Your Own Device (BYOD) programs. This analysis will help in developing a better system. Anu S and Vimala have surveyed various sniffing attacks, namely MAC flooding, ARP poisoning, DHCP attacks, DNS poisoning, password sniffing and presented brief steps for carrying out these attacks [5].

## 2. BACKGROUND

Defense Mechanisms against DDoS attacks, in particular, the sniffing attacks are an emerging research issue. Researchers have proposed and implemented many Intrusion Detection/Prevention Systems (IDS/IPS) to avoid the variants of DDoS attacks. The sniffing is the prominent reason for most of the web attacks and specifically DDoS attacks. A survey on intrusion detection practices that are most commonly used viz., IDS using SVM (Support Vector Machine), GA (Genetic Algorithm), and a cross-layer protocol for DDoS attacks in the sensor nodes was done, and performance of the various classifier was evaluated [6]. A comparative review on service disruptions by DDoS attacks over the internet stated that it is possible to completely stop the DDoS attacks by enhancing the QoS (Quality of Service) of Intrusion forbearance tools [7].
Barth A et al., have done a broad survey on various algorithms for content sniffing used by most popular browsers like Chrome, IE, and so [8]. They have proposed an algorithm to alleviate against the attacks they have encountered through their analysis of content sniffing algorithms with web filters. They were specifically concerned to avoid papers from a look back over themselves. To ensure maximum compatibility and security, they used two principles, namely avoiding privilege escalation and Using prefix disjoint signatures.

Barua et al.,[9] implemented an approach to detect and prevent the content sniffing attacks. They have used the scripting language parsers to analyze the content of the files uploaded through websites and evaluated the results by examining with a suite consisting of both malicious files and benign files. Content sniffing attacks were detected by encrypting the files using Private Key cryptography and by using a file splitter technique [10]. A tag bit generated by the MD5 algorithm is used to check for the modification in the files.

Prerna Arote and Karam Veer Arya proposed a technique using a central server (CS) concept and voting to detect and prevent the ARP poisoning attack. CS will sniff the traffic over the network and sends trap ICMP ping packets to analyze the ICMP replies. By this, it can successfully detect the intruders [11]. To prevent the Central server from attack, they employed the voting process to designate a candid server as CS.

Using the Gratuitous ARP Decision packet System (GDPS), Salim, H et al. proposed a solution to prevent the ARP spoofing attacks [12]. They have compared the performance of the system with normal ARP packets. GDPS system performs the real-time traffic analysis of the interested hosts, and for the received anomalous ARP packets by analysis, a GDP packet is sent for prevention. In this packet, the source address is set to 0.0.0.0 and the mode of transfer is set to uncast.

A Probe packet based technique with enhanced spoof detection engine was used by Poonam Pandey to prevent the ARP cache poisoning attacks [13]. It uses the verifier table and handler algorithm to detect the attacker. ARP-ICMP probe packets were used to detect the ARP spoofing for identifying the legitimate IP-MAC pair.

Attackers launch various attacks like MITM attacks, DoS attacks, cloning attacks through ARP spoofing. Imtiyaz Ahmad lone and Md. Ataullah had done a broad survey on various mitigation techniques against ARP sniffing attacks with their merits and demerits [14]. Anubhi Kulshrestha and Sanjay Kumar Dubey analyzed and reviewed various types of sniffing attacks and concluded that sniffing is the most susceptible attack in browsing. Anubhi Kulshrestha and Sanjay Kumar Dubey had discussed various attack types like password sniffing, content sniffing and phishing attack with the ways to overcome those attacks [15].

Abdul Nasir et al. proposed an intelligent approach for detecting sniffing attacks [16]. This technique is used to detect the active and passive sniffing attack by determining the nature of the network spontaneously. This technique invokes ARP based detection or IP-packet routing based detection if the nature of networking is broadcast or non-broadcast, respectively.

Tasnuva et al., done a systematic analysis of DDoS attacks and its various mitigation technique [17]. They illustrated the various phases of DDoS attacks, identified the features of various DDoS attacks and surveyed the various websites affected by DDoS launch. They had shown that the size of DDoS attacks for the years 2007 to 2016, which was tremendous in 2016 and the frequency of occurrence is also more [18]. They conclude some more new attacks are rising whose features are not predictable. They have concluded stating that DDoS has a greater impact on non-conventional domains like cloud, Big data, smart grids, and IoT.

Arbor WISR annual report states that the security loopholes in IoT devices and botnets have paved the way for the attackers to launch massive attacks based on immense growth in attack size and frequency recorded from 2007 to 2016. It suggests that DDoS mitigation should be the top priority for any organization [19]. 13th NETSCOUT Arbor annual report states that there were more than 100+ attacks recorded by 13% of respondents in 2017. It states 41% of Academic government and enterprises had experienced DDoS attacks[20] while the 14th annual report states 91% of enterprises' network bandwidth were completely saturate by DDoS attacks as its size was 1.7Tbps [21].

## 3. CONTENT SNIFFING AND ARP ATTACKS

Of various sniffing attacks, the attacks concerned with ARP based sniffing and content sniffing was considered for this analysis and the effect of attacks were analyzed using a lattice structure.

### 3.1. Content Sniffing Attacks

According to [21], the content sniffing is a tactic of construing the file types or sniffs the sensitive content from the file. It can be performed on both server side and client side of the web application. Since it is involved in detecting the file formats, content sniffing is also known as Multipurpose Internet Mail Extension (MIME) sniffing [22]. In this method, the attacker gains the information and intentionally uploads files with illicit contents. Though illicit, these files look like genuine one of the MIME types specified. These payloads may cause Phishing attacks, Cross-site scripting (XSS) attacks, and SQL injection attacks, which are the variants of DDoS attacks [1].

#### 3.1.1 Phishing Attacks

A variant of content sniffing is Phishing attacks, which means an e-mail fraud method of sniveling sensitive information. The perpetrator gains access to the information by sending an e-mail which apes the legitimate e-mail and gains the details of the recipients. These emails appear to be from trustworthy sites [23]. Consequently, Ford et al., proposed a tool for detecting the flash advertisements that leads to malicious behaviour in dynamic web content [24]. It works on both static and dynamic nature.

This phishing is not only meant for email, but it can also clone any of the social networking sites like Twitter, Facebook and pretends one to enter the credentials. The phishing can also be through phone, Search engines; Trojan affected hosts, instant messaging, trustworthy sites, individuals or specific company, and so [25, 26]. These types of Phishing attacks are depicted in Figure 2. Several approaches for detecting and preventing phishing attacks are available. Hossain Shahriar and Mohammad Zulkernine had done a comparative analysis of various techniques, websites affected and proposed a new automatic detector Phish Tester [27].

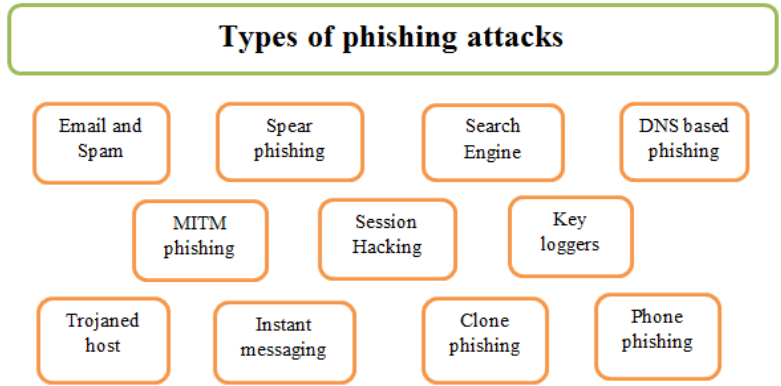

Figure 2. Types of phishing attacks

### 3.1.2 Cross-site scripting (XSS) attacks

XSS vulnerabilities arise when the web pages are not sustained carefully from the attacks. With this loophole, the attackers inject haphazard JavaScript or HTML code which the browsers execute [22]. By this exploitation, the attackers can gain access to the sensitive information entered by users through web pages. By 2011, 60% of websites were vulnerable to these XSS attacks [27], and over 100 million websites (inclusive of famous websites too) were vulnerable to this. The Type- I and Type-II XSS attacks are carried out by persistently storing the inserted script in target servers to attack the victim when here retrieves the stored information from the servers and with the later one the inserted script is reflected off the web server to attack the victim when he clicks the malicious link. Gebre et al., state that content sniffing XSS attacks occur when the content sniffing algorithm of browser and website's differs, the attacker can plinth XSS on page visitor [28] as shown in Figure 3.

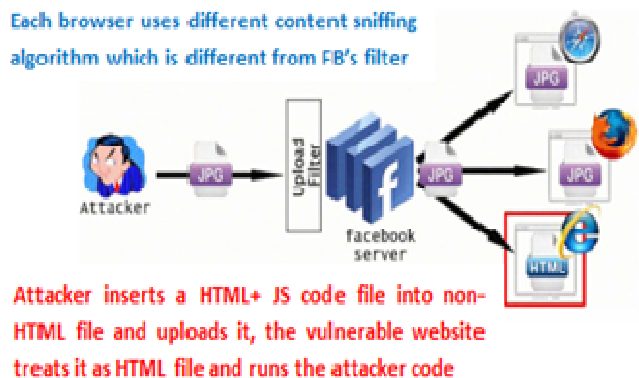

Figure 3. XSS attack by exploiting the mismatch content sniffing algorithms

105

International Journal of Computer Networks & Communications (IJCNC) Vol.11, No.4, July 2019Similar to [28], Barth et al., explained how the attack mounts the XSS vulnerabilities to Wikipedia by embedding HTML code into GIF called as GIF/HTML chameleon, in turn, the browser treats it as HTML file and runs the script [8] and implemented a new technique to detect XSS.

### 3.1.3 SQL Injection Attacks

An SQL injection is the most hazardous on; it stood first of all vulnerabilities whereas XSS at position 3. SQL injection takes place by inserting malicious SQL statements into the SQL query.

> SELECT * FROM Students WHERE Id = 51 or 1=1
>
> SELECT Id, name, Pwd FROM Students WHERE Id = 51 or 1=1

Figure 4. Malicious SQL query

The queries mentioned in Figure 4 give the same results, i.e. both retrieves the student's id, name, password of all students in the database as the condition 1=1 is true always. When the SQL injection query results are not visible to the user, then it is blind SQL injection. To mitigate these attacks, parameterized statements can be used. Sadeghian et al., classify SQL injection attacks under seven major categories which are cause for these attacks [29]. They derived a taxonomy of SQL injection attacks' mitigation mechanisms, which includes detection and prevention techniques adopted by various researchers [14].

### 3.2. ARP Attacks

Address Resolution Protocol is used for mapping the IP address (network-layer address) of a host to its corresponding MAC address (link-layer address). Any host willing to know the MAC address of another host can avail this facility from ARP request-response protocol. ARP has other functionality of which is used for updating a host's changed address in other hosts' ARP table's cache entries, in simpler terms, it is used just for announcements. This announcement is called a gratuitous ARP message. Moreover, it is a stateless protocol which makes entry of all hosts without cross-checking it. Because of this nature, it is vulnerable to attacks such as ARP cache forging which in turn causes the MITM, cloning, DoS and broadcast attacks [14]. The attacks above are active. The passive of ARP attacks is ARP sniffing attack, which will gather sensitive information by observing the packets transmitted.

### 3.2.1. ARP Cache forging

ARP's subtle fault was noticed after it was drafted. It does not enforce any authentication policies for the communicating entities. This paves the way for the attacker to falsify an ARP message with malicious content for poisoning the ARP-cache of the victim's host. This is also termed as cache poisoning. It just broadcast the ARP-request message with IP address to all the hosts in the network to respond with MAC address. So any intruder can make use of this to make a false entry of IP-MAC pair into the victim's table, in turn, the victim will send all the packets to the attacker's MAC address received in the unicast ARP - response message. This attack scenario is depicted in Figure 5. This forms a variant of MITM attacks.

Abhishek Samvedi and Sparsh Owlak had implemented a secure ARP to combat this forging or spoofing [30]. The forged replies specified in the above figure are called as spoofed ARP responses. This type of spoofing can occur at destination, source or victims network traffic.

### 3.2.2. Broadcast attacks

The attacker sends a numerous Ping or ICMP echo request traffic to all IP address in the broadcast list with the spoofed source IP address of the victim. By this the attacker generates required packets with victim's source IP address, consequently sends a sequence packets to any organization with a large number of computers ( to a broadcast address), which in turn is broadcasted to all hosts in the network. Subsequently, the hosts connected to the network will flood the organization with responsibility to the packet sent [12].

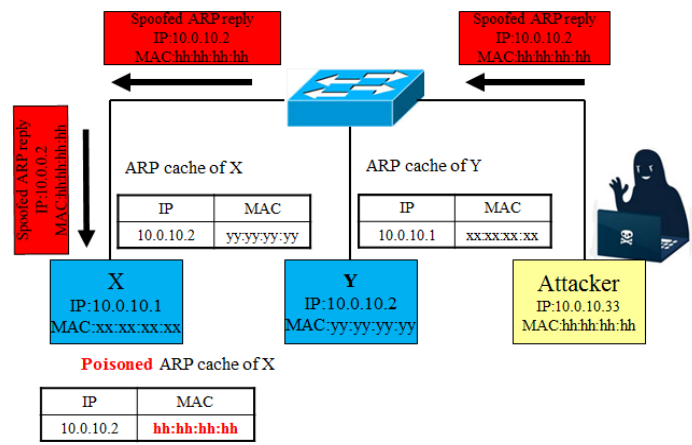

Figure 5. ARP cache Forging

### 3.2.3. Cookie or Session hijacking

In this type of attack, the attacker will sniff the session id of TCP socket connection between two hosts, and by inserting a spoofed packet, he takes control over the connection. In this scenario, it makes the entities to believe the attacker as a trusted server or client. This takes place actively and passively. In the later part, the attacker hijacks the session and records all the traffic passing over it. Various mitigation methods against this attack are specified [14], which includes static entry and some tools.

### 3.2.4. Cloning of IP and MAC addresses

The attacker, with the knowledge of the victim's unique IP and MAC addresses, can use that to his computer. Now the all packets or frames will be received by the attacker [12].

## 4. LATTICE STRUCTURAL ANALYSIS

This section provides an analysis of various attacks caused by content sniffing and ARP attacks, to ensure that sniffing is the prominent activity for DoS or DDoS attacks. A lattice structure can be obtained, proving that these attacks either directly or indirectly cause the DoS or DDoS attacks.

Lattice (L, ≤) is defined as a partially ordered set in which any two elements X and Y has both LUB (lowest-upper-bound) and GLB (greatest-lower-bound). Lattice to be a partially ordered set, it must satisfy the properties viz., reflexive, antisymmetric and transitivity [31].

### 4.1. Purpose of analysis using lattice structure

The lattice structure gives the intricate relationship between the elements participating in the structure. In this context, the attacks taken for analysis are either directly or indirectly connected to each other (most probably indirectly connected). To be in the lattice, it must be a partially ordered set (POSET), where each element must have upper bounds and lower bounds. This characteristic of lattice provides a solution for how these attacks are related.

### 4.2. Lattice Structure

The lattice structure depicted in Figure 6 is a lattice since all the elements have LUB and GLB, satisfies the properties of a partially ordered set. The elements in the lattice structure L are Sniffing (S), Content Sniffing (CS), ARP Sniffing (AS), Broadcast Attacks (BA), Phishing Attacks (PA), ARP Cache Poisoning (CP), DoS (Denial of Service), and DDoS (Distributed Denial of Service).

#### 4.2.1. Assumptions

The Lattice L is the POSET defined s: L= {S, CS, AS, BA, PA, CP, DoS, DDoS}; The relation "≤" is defined over L.

Let X and Y be the element of the lattice and these elements denote the attacks. Then X≤Y (the element X is related to the element Y) if and only if, X causes Y. X and Y are related if X causes Y either directly or indirectly. E.g., Sniffing attack and DDoS are related if and only if Sniffing attack causes DDoS attack either directly or indirectly.

The elements in partially ordered set 'L' on the relation ≤, should satisfy the reflexive, anti-symmetric and transitivity properties defined as:

- Reflexivity: L is reflexive on relation ≤ if and only if every element of L is related to itself  Eg: X ≤ X and it should hold for all the elements of L
- Anti-symmetric: L is anti-symmetric on relation ≤ if and only if  for every element l1 and l2 in L, l1 ≤ l2  and  l2≤l1
- Transitivity: L on ≤ is transitive when l1∈ L, l2 ∈ L and l3 ∈L, if l1 ≤ l2, l2 ≤ l3 the l1 ≤ l3 holds.

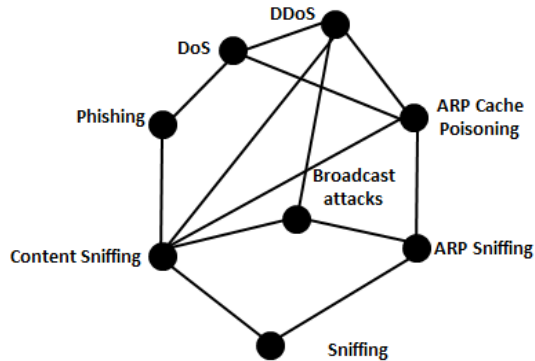

Figure 6. Structure of Lattice (L,≤ )

These are the elements of POSET under relation "≤" as it is undeniable from the survey made that they satisfy the reflexive, anti-symmetric and transitivity properties.

*CS And AS*
LUB (CS, AS):
➔ CS + AS
➔ BA

Here CS ≤ BA, because only by sniffing the contents of the IP or MAC details BA will happen. Also AS ≤ BA, since the content sniffed are from ARP table. CS and AS have other UB, but the least is BA.

GLB (CS, AS):
➔ CS * AS
➔ S

Here S is the greatest lower bound of CS and AS because it is a process which causes both CS and AS

*P and CP*
LUB (P, CP):
➔ P + CP
➔ DoS

Here P ≤ DoS, as one of the types of phishing, namely phone phishing which makes un-wanted calls or messages to the customers impersonating the actual service provider by fake calls.

Moreover, ARP cache poisoning makes false entries in the victim's table, consequently makes the victim deny his services to legitimate users. Therefore, CP ≤ DoS are related. P and CP have other UB, but the least is DoS.

GLB (P, CP):
➔ P * CP
➔ CS

Here CS is the greatest lower bound of P and CP because CS is the preliminary activity for these two attacks.

*CP and DoS*
LUB (CP, DoS):
➔ CP + DoS
➔ DoS

Here CP ≤ DoS, as one of the major reason for the variant of DoS attacks by ARP spoofing is CP. As DoS is self-exhaustive, DoS ≤ DoS. Therefore, the least upper bound of CP and DoS is DoS.

GLB (CP, DoS):

109

➔ CP * DoS
➔ CS

Here CS is the greatest lower bound of CP and DoS because CS is the preliminary activity for these two attacks.

*DoS and DDoS*
LUB (DoS, DDoS):
➔ DoS + DDoS
➔ DDoS

Here DoS ≤ DDoS, because DDoS attacks are the most prominent variants of DoS attacks. Consequently, DDoS ≤ DoS since DDoS compromises the individual nodes to act as agents, thus denying their services to their legitimate users.

GLB (DoS, DDoS):
➔ DDoS * DoS
➔ S

Here S is the greatest lower bound of DDoS and DoS because S is the preliminary activity for these two attacks.

Hence it can be concluded that this Lattice (L, ≤) is also a bounded lattice with the bounds are [S, DDoS]. Because S is the process without which these attacks have no role to play as well DDoS is the making possible greatest devastating effects of all the elements or attacks or activities in the POSET.

### 4.3. Application of lattice structure

This lattice model can be used to perform the following:

• To validate any Solution for DoS attacks by checking whether the system mitigates all the attack elements in the lattice.
• Helps to test the mitigation system against possible attack scenarios like: "If Phishing attack is detected by a mitigation technique, will this technique also detect DoS attack?"
• By using this, we can compile the relation between various attacks detected or prevented and can trace out the cause for each attack type, devastated victim resources.

The following Cross-Layer Consistency Checking (CLCC) technique by B.Prabadevi and N.Jeyanthi can mitigate ARP sniffing attacks [32]. CLCC is evaluated using the lattice model. The system consists of the following components: Packet analysis, ARP cache updating, fake list updation and generation of broadcast Alert Message. The packet analysis component determines the correct ARP packet depending on the opcode, if the opcode is 1 it is ARP Request, if 2 it is ARP Reply, if 25 it is broadcast Alert message and if 26 it is a unicast alert message to the router.

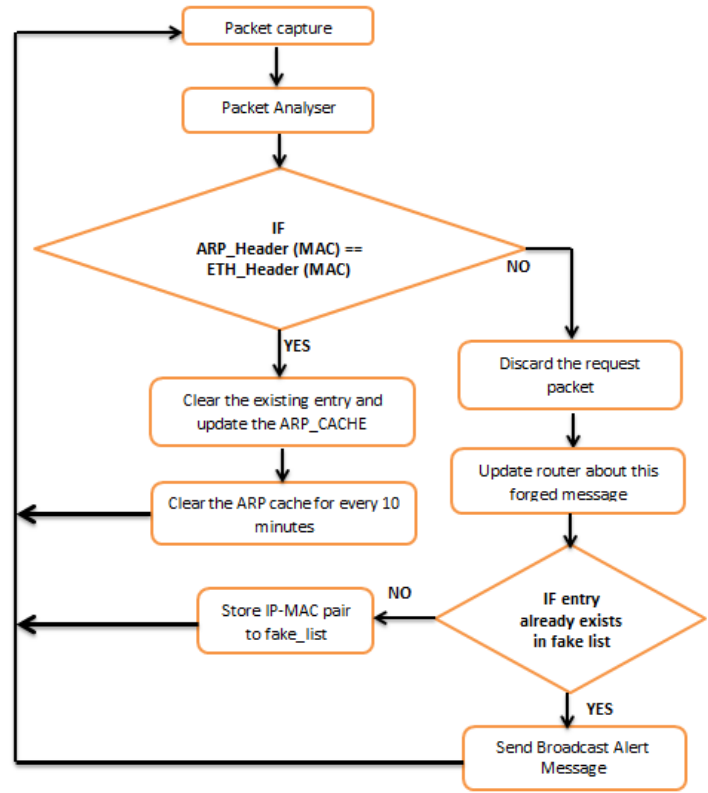

Figure 7. A mitigation technique for ARP Cache poisoning attack

The system illustrated in Figure 7 works as follows:

**At Router**
**Assumptions:** Router has received a unicast alert message
**Variables:** arp_IPA, arp_MACA, arp_IPR, arp_MACR, Eth_MACA, Eth_MACR
**Step: 1**
**If**(opcode==26)**Then //** Unicast Alert Message
    **If** (arp_IPR==IPR) **Then**
        **If** (arp_MACA==Eth_MACA)**Then**
            **If** (arp_IPA && arp_MACA exists in Router cache) **Then**
                Accept Alert;
            **Else**
                Add (arp_IPA, arp_MACA) to fake list;
                Ignore Alert;
        **Else**
            Ignore Packet and add to fake list;

**Working Procedure:**
**Variables:** arp_IPA, arp_IPD, arp_MACA, arp_MACD, Eth_MACA,,Eth_MACD.
When host A wants to know host D's MAC :
// sender→ arp_IPA, arp_MACA and Eth_MACA
// target→ arp_IPD, arp_MACD and Eth_MACD

**Step1:**
send ARP_Request_Message (ARP Header, Ethernet Header);  //opcode=1
**Step 2:**      /*At D*/
   packet analysis();// checking for opcode==1
   Decode_the _packet();

**Step 3:**
**if** (arp_MACA==Eth_MACA) **then**
   Update_arp_cache();
   send_ARP_Reply_Message (ARP Header, Ethernet Header); //opcode=2
**else**
   Store the forged IP- MAC_address into fake_list;
      send alert message to the_Router or gateway;//opcode=26
   if(arp_IPA and arp_MACA already exists in fake_list)
      send Broadcast_Alert_Message;//opcode=25

**Step 4:**       /*At A*/
// sender→ arp_IPD, arp_MACD and Eth_MACD
//target→ arp_IPA,arp_MACA and Eth_MACA
   **Repeat Step 2 and Step 3;**
**Step 5: Clear the ARP-Cache for every 10 minutes**

**4.4.1. Evaluation of the proposed Algorithm**

The proposed method is evaluated by computing the complexity of the algorithm. The algorithm performs a cross-layer check(ARP and Ethernet layers) before processing ARP reply or Request packets done in fixed number steps: O(1) since the values are constant. It has two new messages apart from two messages in the traditional method and maintains a fake list which is updated for each forged entry found. This entry is scanned once for each forged entry and if a match found the broadcast message is generated. Consider the entries in the fake list is initially empty but in worst case 'k' entries. Then this scan incurs O (k) and a sequential scan for each ARP Reply – Request message incurs O (j) if it has j entries. Since the cache is cleared for every 10 minutes, this complexity may vary depending on the hosts in the network. O (1) for each comparison done before generating reply and updating fake entries. On the worst case, the total complexity will not be more than O (j) + O (k) +scanning cost. In the best case, it may be lesser than several hosts in the network.

Table 1 compares the proposed algorithm with existing techniques RFC 826 [33], SARP [34], TARP [35], EARP [36] and Central Server Approach [37]. Though the proposed system performs

International Journal of Computer Networks & Communications (IJCNC) Vol.11, No.4, July 2019cross-layer checking, non-cryptographic still it incurs some of the attacks caused through IP spoofing and router poisoning. The proposed can withstand ARP cache poisoning attacks like MiTM, DoS, host impersonation, cloning to some extent but attacks caused by ARP scanning and forging still pertains. The problems with gratuitous ARP requests yet to be covered.

The proposed algorithm is simulated as follows: Three nodes viz., A with IP-MAC pair 192.169.1.10-00:5:79:66:68:01, B with IP-MAC pair 192.169.1.11-00:5:79:66:68:02 and C with IP-MAC pair 192.169.1.12 -00:05:79:66:68:03.

Table 1. Comparison with Existing solutions

| Features | RFC826 | SARP | TARP | EARP | GARP | Central Server | Proposed |
|---|---|---|---|---|---|---|---|
| Cross Layer Inspection | No | No | No | No | No | No | Yes |
| ARP Stateful | No | Yes | Yes | Yes | Yes | Yes | Yes |
| ARP storm Prevention | No | No | Partial* | Yes | Yes | Yes | Partial |
| Static-S and Dynamic-D entries | S&D | D | D | S&D | S&D | S&D | S&D |
| Cryptographic | No | Yes | Yes | No | Yes | Yes | No |
| * leads to ticket flooding attack | | | | | | | |

The proposed algorithm detects the following type of packets:

    PKT#1. ($MAC_{VAL}$, $IP_{INV}$) in ARP Request(Destination)
    PKT#2. ($MAC_{INV}$, $IP_{VAL}$) in ARP Request(Source)
    PKT#3. ($MAC_{VAL}$, $IP_{INV}$) [This situation may even happen when the hosts are assigned IP dynamically]
    PKT#4. ($MAC_{INV}$, $IP_{VAL}$) in ARP Reply(Destination)
    PKT#5. ($MAC_{VAL}$, $IP_{INV}$) in ARP Reply (Source)
    PKT#6. ($MAC_{VAL}$, $IP_{INV}$) in ARP Reply during the dynamic assignment
    PKT#7. ($MAC_{INV}$, $IP_{VAL}$) in Broadcast Alert Message
    PKT#8. Null MAC Address in Broadcast Alert Message
    PKT#9. ($MAC_{INV}$, $IP_{VAL}$) in Unicast Alert message (Source)
    PKT#10. ($MAC_{VAL}$, $IP_{INV}$) in Unicast Alert message (Destination)
    PKT#11. ($MAC_{VAL}$, $IP_{INV}$) in Unicast Alert Message (Source)

$MAC_{INV}$ → Invalid MAC Address
$IP_{INV}$ → Invalid IP Address
$MAC_{VAL}$ → Valid MAC Address
$IP_{VAL}$ → Valid IP Address
NULL → Null MAC address which is non-existent

Table 2 below provides the details about the possible packets that the proposed system detects. These packets, when detected, can avoid DoS, MiTM, and host impersonation to some extent.

Though the proposed system can able to detect these packets, still the same problem persists with this ARP viz., a host will still receive a reply for an unsent request. The fake list table used here is never cleared so that the entries may provide expired details, and it may leave the network in chaos. The fake list table may grow larger in an attack situation, which may consume larger storage. This situation may lead to some new attacks as well. The replay attacks are not tracked up, which can be avoided to some extent only.

However, the packets PKT#1, PKT#7 and PKT#8 can be detected only if the request or reply received before cache clearance or it has a static entry of the devices in the network.

However, the packets PKT#1, PKT#7 and PKT#8 can be detected only if the request or reply received before cache clearance or it has a static entry of the devices in the network.

Table 2. Packets that proposed algorithm detects

| Packet No | Detection Situation | Source / Destination | Packet type | The feature that does this | Sample |
|---|---|---|---|---|---|
| *AT B, Assume Host A wants to communicate with host B* | | | | | |
| PKT#1 | Only if the victim received the request before cache clearance of its previous entry | Destination host | ARP Request | Cache entry checking and Fake list entry | **ARP_EntryA in B:** 192.169.1.10 **ARP_IPA:** 192.169.1.17 (100% detection if entry is in the fake list even after cache clearance) |
| PKT#2 | When ARP-MAC does not match with ETH_MAC | Source host | ARP Request | Cross-Layer checking | **ARP_MACA:** 00:5:79:66:68:12 **ETH_MACA** 00:5:79:66:63:01 |
| PKT#3 | Victim detects its IP is wrong | Source host | ARP Request | ARP cache entry checking | **ARP_IPB:** 192.169.1.11 **ARP_Source:** 192.169.3.23 |
| *AT A, Assume Host B in the process of Accepting ARP Requests from A* | | | | | |
| PKT#4 | When Eth-MAC and ARP_MAC does not match | Destination host | ARP Reply | Cross-layer checking | **ARP_MACA:** 00:5:79:66:68:12 **ETH_MACA** 00:5:79:66:63:01 |
| PKT#5 | If the victim received the reply with spoofed IP | Destination host | ARP Reply | Cache entry checking and Fake list entry | **ARP_IPA:** 192.169.1.10 **ARP_Source:** 192.169.3.26 |

International Journal of Computer Networks & Communications (IJCNC) Vol.11, No.4, July 2019

| PKT#6 | Victim detects its IP is wrong | Source host | ARP Reply | On reception of ARP packet and Entry checking | **B's IP:** 192.169.1.11 **ARP_IPB:** 192.169.1.18 |
|---|---|---|---|---|---|
| **Assume A sends a Broadcast message** | | | | | |
| PKT#7 | All the host in the network can detect from its static ARP entry | Source host | Broadcast Alert Message | Cross-layer checking | **ARP_MACA:** 00:5:79:66:68:AF **ETH_MACA** 00:5:79:66:63:01 |
| PKT#8 | All the host in the network can detect from its static ARP entry | Source host | Broadcast Alert Message | Cross-layer checking | **ARP_MACA:** 00:00:00:00:00:00 **ETH_MACA** 00:5:79:66:63:01 |
| **An Alert message from host A to Router (At Router)** | | | | | |
| PKT#9 | When the ARP_MAC and ETH_MAC does not match | Source host | Unicast Alert Message | Cross-Layer checking | **ARP_MACA:** 00:06:80:99:80:00 **ETH_MACA** 00:5:79:66:63:01 |
| PKT#10 | When the Router's IP does not match | Destination host | Unicast Alert Message | On message reception | **ARP_IP of Router:** 10.10.1.1 **Router's IP:** 10.10.1.0 |
| PKT#11 | When ARP_IPA does not match with Router's ARP cache | Source host | Unicast Alert Message | ARP cache checking | **ARP_IPA:** 192.169.1.17 **ARP_IPA in router's cache** 192.169.1.10 |

The traditional ARP system is vulnerable to the following type of attacks: MitM attack, DoS, hijacking connections through sniffing, making clones or impersonating the hosts, spoofing IP and MAC addresses, IP conflict attack, ARP request/reply flooding and replay attack.

Of the attacks specified the proposed algorithm can mitigate following types of attacks as follows: The man-in-the-middle attack that is caused by IP-spoofing and MAC-spoofing can be avoided. The algorithm encounters these two cases by cross-layer checking and fake-list updation. But IP-spoofing may still prevail in case of dynamic IP configuration, as fake list entry clearance is not done. This eventually leads to denial of service attacks, which will be consuming all the network resources of the victim. In turn, the legitimate users of the network are starved.

International Journal of Computer Networks & Communications (IJCNC) Vol.11, No.4, July 2019

The tremendous form of DoS is distributed-DoS, where they can compromise the network resources faster than the DoS by making various compromised hosts to flood the network with unsolicited messages.

The cloning of host will be mitigated by cross-layer checking, as hosts in the LAN can communicate via MAC addresses only. Though MAC spoofing will be mitigated, IP spoofing may still exist. Since the proposed algorithm does not bother about timely delivery of ARP messages replay attack will be sustained by which attack tries to bombard the network by sending the same packets again and again until the network is frustrating.

The graph in Figure 16 depicts how the packet detection rate in comparison with other systems. The proposed algorithm is compared with ARP [33], TARP [35], EARP [36], GARP [37] and Centralized Approach [38]. It's been clear that the proposed algorithm performs better than all the techniques. The following formula obtains the packet detection rate:

**PDR(%) = #APD / #TMP** …………………..    Eqn.1

Where PDR➔Packet Detection Rate; #APD➔Number of Abnormal packets detected #TMP➔ Total number of abnormal packets sent. The PDR of CLCC and RFC 826 is given in Table 3.

Table 3. Comparison of CLCC with RFC826

| ARP Techniques | Total number of Packets injected | | No. of malicious packets detected | Detection Rate |
|---|---|---|---|---|
| | *Normal Packets | **Abnormal Packets | | |
| RFC 826 | 100 | 1155 | 115 | 12 |
| CLCC | 100 | 1155 | 892 | 77 |
| *Normal Packets➔Normal ARP Request Reply packets  **Abnormal Packets➔ PKT#1 to PKT#11 | | | | |

The graph is obtained by analysing the systems with various type of packets listed in Table 2.

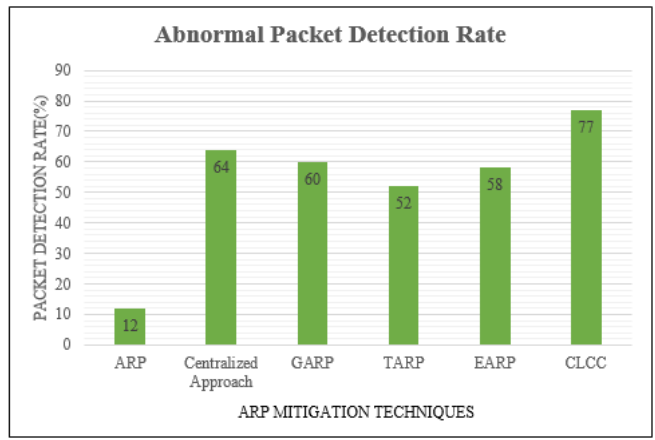

Figure 8. Graphical analysis of proposed with existing systems

## 5. CONCLUSIONS

A broad study on content sniffing and ARP sniffing attacks were conducted. A lattice structure obtained guarantees that sniffing activity causes DoS or DDoS attacks. This lattice model will help to ensure that any detection mechanism for a DDoS attack can combat all the attacks mentioned in the model. Also, the proposed ARP mitigation techniques prove that ARP - sniffing causes the DoS or DDoS attack, and this system mitigates it by cross-layer checks and fake list updation. The mitigation technique proves that ARP sniffing causes the DoS or DDoS attack, and this system mitigates it by cross-layer checks and fake list updation. This technique avoids host impersonation, DoS attacks and MiTM attacks caused by ARP cache poisoning. Since it clears the cache every 10mins, the ARP table has to be updated each time a new host communicates in the network. Though this can be a loophole for the attackers, the checking of Ethernet header information with ARP header information, they will go vain without any information being retrieved. This will be an authenticated method which avoids most of the attacks by cache poisoning.